# The wall shear rate in non-Newtonian turbulent pipe flow

Trinh, Khanh Tuoc

K.T.Trinh@massey.ac.nz

**Abstract**

This paper presents a method for calculating the wall shear rate in pipe turbulent flow. It collapses adequately the data measured in laminar flow and turbulent flow into a single flow curve and gives the basis for the design of turbulent flow viscometers.

## 1 Introduction

All turbulent non-Newtonian turbulent transport phenomena have been studied using rheological parameters determined separately with rheometers operating in laminar flow. In the general case, the flow curve, a plot of shear stress $\tau$ against shear rate $\dot{\gamma}$, is not linear, even on a log-log plot. Thus the experimental flow curve cannot be extrapolated with confidence beyond the range of the measured data and the value of $n = \log\tau/\log\dot{\gamma}$ must be obtained at the shear rate/stress prevalent in the turbulent flow. For example Metzner (Clapp, 1961) criticised the validation of Clapp's correlation of turbulent heat transfer in pseudoplastic fluids because the maximum shear stress achieved in the rheological measurements did not cover the range covered in the turbulent transfer experiments.

The next question that we should immediately address is: what evidence exists to prove that rheological parameters determined in laminar flow are still valid in turbulent flow? Without this proof, there is no real solid basis for studies of non-Newtonian transport. Surprisingly there is still no publication on this issue. This paper reports work that I performed some time ago on this problem (Trinh, 1969).

## 2 The wall shear stress in laminar pipe flow

We analyse pipe flow as a case study but the methodology also applies to other rheometer configurations.

The shear rate cannot be measured directly but can be calculated from the average flow velocity through the well known Mooney-Rabinowitsch equation (Skelland, 1967)

$$3\tau_w^2 \frac{Q}{\pi R^3} + \tau_w^3 \left( \frac{d \frac{Q}{\pi R^3}}{d\tau_w} \right) = \tau_w^2 f(\tau_w) = \dot{\gamma}_w \tau_w^2 \quad (1)$$

$$\dot{\gamma}_w = f(\tau_w) = \frac{3Q}{\pi R^3} + \tau_w \left( \frac{d \frac{Q}{\pi R^3}}{d\tau_w} \right) \quad (2)$$

The quantity

$$\Gamma = \frac{4Q}{\pi R^3} = \frac{8V}{D} \quad (3)$$

has been called "average shear rate" by Severs and Austin (1954), "nominal shear rate at the pipe wall" by Symonds, Rosenthal, and Shaw (1955), "apparent shear rate at the pipe wall" by McMillen (1948), and "flow function" by Bowen (1961) but a more suitable name is *nominal wall shear rate* since it is the value returned by most commercial computerised viscometers and corresponds to the wall shear rate calculated with a Newtonian fluid formula.

We now introduce the symbol

$$n' = \frac{d \ln(\tau_w)}{d \ln(\Gamma)} \quad (4)$$

$$\dot{\gamma}_w = \left( \frac{3n'+1}{4n'} \right) \Gamma = \left( \frac{3n'+1}{4n'} \right) \left( \frac{8V}{D} \right) \quad (5)$$

The wall shear stress then becomes

$$\tau_w = K' \left( \frac{8V}{D} \right)^{n'} = K \left( \frac{3n'+1}{4n'} \right)^{n'} \left( \frac{8V}{D} \right)^{n'} \quad (6)$$

Equations (5) and (6) are used to determine the flow curve for fluids studied in rheometry.

## 3    The wall shear rate in turbulent pipe flow

The friction factor in purely viscous non-Newtonian turbulent flow has been correlated by Dodge and Metzner (1959) with an extension of the Blasius (1913) correlation for Newtonian fluids

$$f = \frac{2\tau_w}{\rho V^2} = \frac{\alpha}{\text{Re}_g^\beta} \tag{7}$$

where

$$\text{Re}_g = \frac{D^{n'} V^{2-n'} \rho}{K' 8^{n'-1}} \tag{8}$$

is called the Metzner-Reed (1955) generalised Reynolds number. The coefficients $\alpha$ and $\beta$ were fitted empirically. These coefficients can be determined theoretically for a power law fluid as (Trinh, 2010)

$$\alpha = \frac{2^{\frac{n+4}{(3n+1)}} .817^{-\frac{7n}{(3n+1)}} 45.82^{\frac{-6n}{3n+1}} (n+1) 2.08^{\frac{7n}{3n+1}}}{\left(\frac{3n+1}{4n}\right)^{\frac{n}{3n+1}}} \tag{9}$$

or

$$\alpha = \left(0.079 \frac{n+1}{2}\right)^{\frac{4n}{3n+1}} (2)^{\frac{5(1-n)}{3n+1}} \left(\frac{4n}{3n+1}\right)^{\frac{n}{3n+1}} \left(\frac{2}{n+1}\right)^{\frac{n-1}{3n+1}} \tag{10}$$

$$\beta = \frac{1}{3n+1} \tag{11}$$

Equations (9) and (10) are alternative formulations that both correlate experimental data adequately (Trinh, 2010). We generalise the application of equations (9), (10) and (11) to other non-Newtonian fluids by adopting a technique used by Metzner and his colleagues. We divide the flow curve

$$\tau_w = K \dot{\gamma}_w^n \tag{12}$$

into small sections that can be treated as straight lines on a log-log plot. Then $n$ is replaced with $n'$ in equations (9) to (10). The shear rate can be expressed as

$$\dot{\gamma}_w = \left(\frac{\tau_w}{K}\right)^{1/n'} \quad (13)$$

$$\tau_w = \frac{\alpha K'^{\beta} 8^{\beta(n'-1)} V^{2-\beta(2-n')}}{\left(2D^{n'\beta}\rho^{\beta-1}\right)} \quad (14)$$

With these considerations, equation (7) can be rearranged as (Trinh, 1969)

$$\dot{\gamma}_w = \left[\frac{\alpha V^{2-2\beta+n'\beta} \rho^{1-\beta} K^{\beta-1} 8^{\beta(1-n')} \left(\frac{3n'+1}{4n'}\right)^{\beta n'}}{2D^{\beta n'}}\right]^{1/n'} \quad (15)$$

## 4    Comparison with experimental data and discussion

We can use the widely quoted data of Dodge and Metzner (1959) to test this theory. Most specifically we will use runs 7a, 7b and 7c using 3% Carbopol flowing in half inch, one inch and two inch brass pipes.

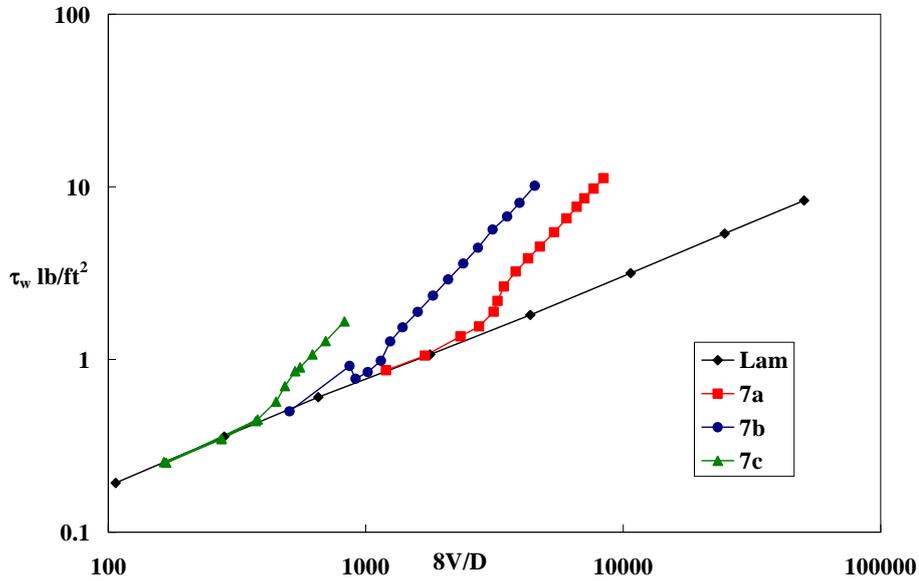

Figure 1. Plots of $\tau_w$ vs. $8V/D$ for 3% Carbopol. Data of Dodge (1959)

Figure 1 shows that the log-log plots of $\tau_w$ vs. $8V/D$ results in different lines for laminar and turbulent flows as first pointed out by Bowen (1961). Fitting equation (6) to the laminar flow data gives $n' = .62$ and $K' = 0.009\, lb.s^{n'}/ft^2$ (Dodge op.cit.).

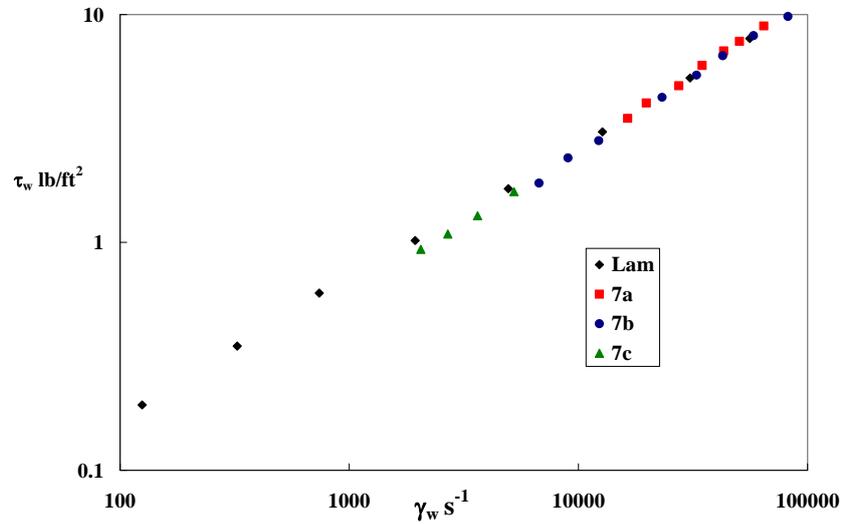

Figure 2  Flow curve for 3% Carbopol using both laminar and turbulent data of Dodge (1959).

Another example is shown with 2% Carbopol in Figures 3 and 4.

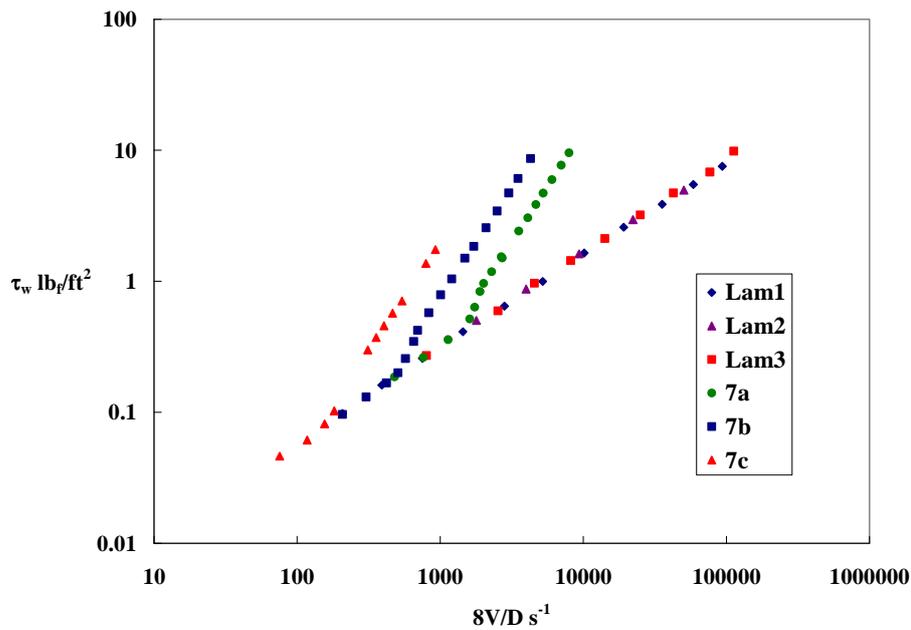

Figure 3. Plots of $\tau_w$ vs. $8V/D$ for 2% Carbopol. Data of Dodge (1959)

The present verification confirms that rheological parameters measured under laminar flow can indeed be used for the study of turbulent phenomena. It also provides the basis for the design of turbulent flow rheometers. We need to keep in mind, however, that unlike laminar flow, turbulent flow is affected by viscoelastic properties of non-Newtonian fluids (Trinh, 1969, 2009) and equation (15) must be modified for viscoelastic fluids.

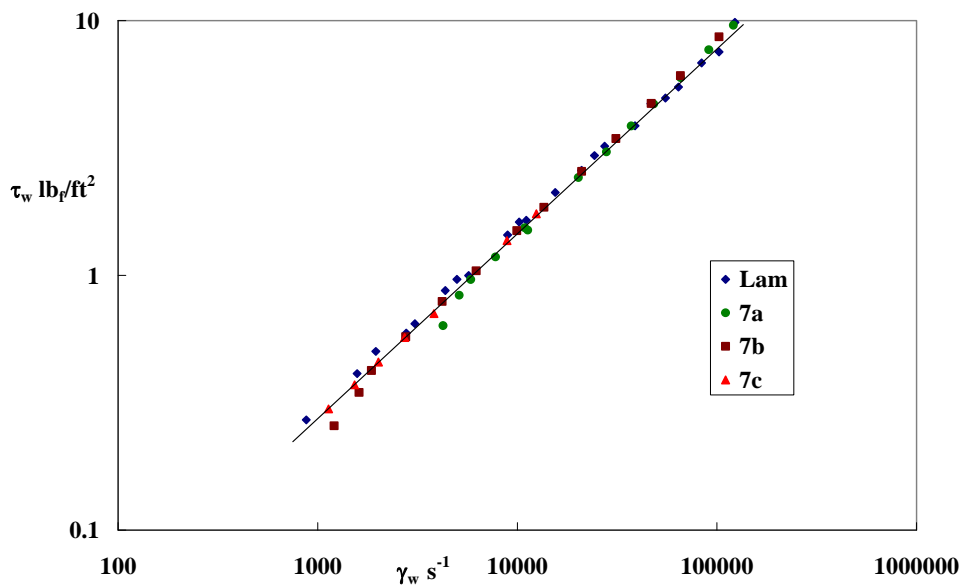

Figure 4. Flow curve of 2% Carbopol using both laminar and turbulent data. Data of Dodge (1959).

## 5    Conclusion

A correlation has been derived for the wall shear stress in turbulent purely viscous non-Newtonian fluids. It confirms that rheological parameters measured in viscometers that operate in laminar flow are still valid for studies of turbulent transport phenomena.